# TeraScale SneakerNet:
# Using Inexpensive Disks for Backup, Archiving, and Data Exchange.

Jim Gray,
Wyman Chong,
Tom Barclay,
Alex Szalay,
Jan Vandenberg

# TeraScale SneakerNet:
## Using Inexpensive Disks for Backup, Archiving, and Data Exchange.
Jim Gray, Wyman Chong, Tom Barclay, Alex Szalay, Jan Vandenberg
May 2002

## The Problem: How do we exchange Terabyte datasets?

We want to send 10TB (terabytes, which is ten million megabytes) to many of our colleagues so that they can have a private copy of the Sloan Digital Sky Survey (http://skyserver.sdss.org/) or of the TerraServer data (http://TerraServer.net/).

**Table 1**. Time to send a TB
| | |
|---|---|
| 1 Gbps | 3 hours |
| 100 Mbps | 1 day |
| 1 Mbps | 3 months |

The data interchange problem is the extreme form of backup and archiving. In backup or archiving you are mailing the data to yourself in the future (for restore). So, a data interchange solution is likely a good solution for backup/restore and for archival storage.

What is the best way to move a terabyte from place to place? The Next Generation Internet (NGI) promised gigabit per second bandwidth desktop-to-desktop by the year 2000. So, if you have the Next Generation Internet, then this transfer is just 8 trillion bits, or about 8,000 seconds – a few hours wait. Unfortunately, most of us are still waiting for the Next Generation Internet -- we measure bandwidth among our colleagues at between 1 megabits per second (mbps) and 100 mbps. So, it takes us days or months to move a terabyte from place to place using the Last Generation Internet[1].

**Table 2**: The raw price of bandwidth, the true price is more than twice this when staff, router, and support costs are included. Raw prices are higher in some parts of the world.

| Context | Speed Mbps | Rent $/month | Raw $/Mbps | Raw $/TB sent | Time/TB days |
|---|---|---|---|---|---|
| home phone | 0.04 | 40 | 1,000 | 3,086 | 6 years |
| home DSL | 0.6 | 70 | 117 | 360 | 5 months |
| T1 | 1.5 | 1,200 | 800 | 2,469 | 2 months |
| T3 | 43 | 28,000 | 651 | 2,010 | 2 days |
| OC3 | 155 | 49,000 | 316 | 976 | 14 hours |
| 100 Mpbs | 100 | | | | 1 day |
| Gbps | 1000 | | | | 2.2 hours |
| OC192 | 9600 | 1,920,000 | 200 | 617 | 14 minutes |

Most of us do not pay the phone bill for our Internet connection; our employer pays the bill. But bandwidth is expensive. Current bandwidth prices are given in Table 2[2]. Huge consumers pay about 200$/Mbps/mo while most people pay much higher tariffs. Interestingly, incoming DSL is the least expensive link – but it is expensive in the outgoing direction.

## Sneaker Net

In recent memory, people exchanged data via floppy disks – this was called *sneaker net*. Today, sneaker net has evolved to CDs. You can write a CD in ten minutes, mail it overnight for a few dollars, and then read it in an hour. The elapsed time is something like 24 hours and the bandwidth is about a gigabyte per day. So, CD sneaker net runs at about 70kbps, costs about 1600 CDs for a terabyte (== 250$ media cost), and takes about a week to write, ship, and read the terabyte. This is cheaper than a home phone and faster than a DSL or a T1 line, but it is not a solution to our problem. There are rumors of 100GB writeable CDs with respectable (10MBps) data rates. When they arrive, they may be an alternative. But, what can we do till then?

---

[1] We sometimes wonder where the 500 M$ for the NGI went.
[2] These numbers are based on anecdotal evidence (Microsoft, Compaq,…) for the OC192 and home, and on http://www.broadband-internet-provider.com/research-information.htm for the T1, T3, OC3 prices.

**Table 3**: The relative cost of sneaker-net, using various media. The analysis assumes 6MBps tape, 10MBps CD/DVD and robots at each end to handle the media. Note that the price of media is less than the fixed robot cost.

|  | Media | Robot$ | Media$ | TB read + write time | ship time | TotalTime /TB | Mbps | Cost (10 TB) | $/TB shipped |
|---|---|---|---|---|---|---|---|---|---|
| **CD** | 1500 | 2x800 | 240 | 60 hrs | 24 hrs | 6 days | 28 | $2,080 | $208 |
| **DVD** | 200 | 2x8000 | 400 | 60 hrs | 24 hrs | 6 days | 28 | $20,000 | $2,000 |
| **Tape** | 25 | 2x15,000 | 1000 | 92 hrs | 24 hrs | 5 days | 18 | $31,000 | $3,100 |
| **DiskBrick** | 7 | 1,000 | 1,400 | 19 hrs | 24 hrs | 2 days | 52 | $2,600 | $260 |

## The case against tape

Today the *best practices* answer to data exchange is "Send tape of course!" But, 1 TB of tape costs more than 1,000$ for the media, a modest tape robot costs 15,000$, and there are serious standards problems: "Do you do DLT, AIT, LTO, or Exabyte?" To add to our woes, tape is slow to write (at 6 MBps it takes 46 hours to write a terabyte). But the fundamental problem is that tape is an unreliable data interchange medium. One stimulus for this work was our failed effort to move the 5 TB teraserver from one data center to another – we could write the tapes (slowly) but we could never read them all. In the end we wrote to a portable disk array and moved the bytes that way.

Suppose we use tapes to move our terabytes. If we use the tapes 10 times, then our cost (you and me) to exchange a TB is 30 K$ for two tape robots and 1K$ for the media. Ignoring labor costs, it is 31k$ total or 3k$ per TB. This is comparable in cost, but much slower and more labor intensive than us each renting part of a T3 line.

## Disk Bricks: Portable Terabytes

We believe that, until the NGI arrives, the best way to send terabytes around is to send computers housing inexpensive disks – *StorageBricks*. This is *terascale-sneakernet*.

We began sending raw disks to one another, but that has the nasty problem that disks do not plug right into the network. So the recipient had to have exactly the right kind of disk reader (an ATA path that could read an NTFS file system with an SQL database). At a minimum, this excludes our Macintosh, Solaris, MVS, and Linux colleagues. Even within the select group that can read our favored disk format, we had many problems about master-slave settings on the ATA bus, about recognizing dynamic disks, and innumerable other details. So, sending magnetic disks around is problematic.

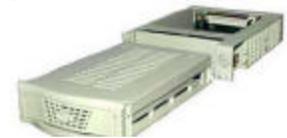

**Figure 1**: We began by exchanging raw disks (or disks in hot pluggable carriers pictured here. But disk format problems (what file system?, …) and low level disk configuration master-slave, plug-and-play,… make this problematic.

Still, for fidelity to the argument, we can read and write disks at about 25MBps (average) and they can be written in parallel at about 25MBps (see table 5). So we can write a 1 terabyte array in about 4 hours, ship the 7 disks overnight for about 50$, and then read them in 4 hours. The disks themselves cost about 1,300$, but they are reusable. If we use them ten times this is a 190$ per shipped terabyte, and the data rate is about 100Mbps. This beats the price of anything in Table 2, and is faster than anything but the (yet-to arrive) NGI.

*To solve the plug-and-play problem of data interchange, we propose to ship (or archive) whole computers*. Why not send file servers and database servers around? These boxes have a power plug that works on 110 and 220 volts, an RJ45 network plug that works on 10-100-1000 Ethernet, and a huge stack of software standards that gives plug-and-play (dhcp, http, NFS, CIFS, ODBC,…). So, the data distribution goes as follows:
(1) Publisher writes the data on the storage brick,
(2) unplugs it and ships it to the subscriber via express mail.
(3) Subscriber plugs the brick in and it joins his network,
(4) The subscriber then reads the data from the brick.
This sounds great, but there are a few details that scare people.
- **Security**: Can you spell firewall? This foreign computer could have terrible viruses. How do you prevent it from hurting anything?
- **Performance**: Does this really work? Can I really get 80 MBps from the brick?
- **Cost**: Aren't computers expensive?

The cost issue is the easiest to dispose of, so let's do that first.

## Storage Brick Cost:

Table 4 is the bill-of-materials for a 1 TB storage brick – about $2,500 in May 2002. The prices will have dropped by the time you read this. The price per terabyte dropped five fold in the last 2 years and will likely decline at least 60%/year for several years to come.

The resulting storage brick, a 3GT (a Ghz processor, a GB of ram, a Gbps Ethernet, and a Terabyte of disk), is shown in Figure 2. The expensive cabinet gives power and cooling for the 7 disks. The Motherboard's raid controller and 4 ATA/133 controllers support the seven 153GB disks configured as a 1.1 TB RAID0 stripe set (using host-based striping in our case). Gbps Ethernet allows the brick to serve data at 35 MBps (3x the 12 MBps limit of 100Mbps Ethernet.)

**Table 4:** The price list for a Terabyte Brick in a 3G host (GHz processor, GB of memory, and Gbps Ethernet)
http://pricewatch.com/

| Item | Price |
|---|---|
| Cabinet (Lian LiPC-68 USG 12 bay case) | 138 |
| Power Supply (Enermax EG465AX-VD 431W) | 117 |
| Motherboard (Abit KX7A-RAID KT266A) | 108 |
| Cpu (AMD 2GHz Athlon XP 1800+) | 110 |
| 1 GB Memory (2x512MB PC2100 266MHz DDR) | 120 |
| 1 TB Disks (7xMaxtor EIDE 153GB ATA/133 5400RPM) | 1,281 |
| Gbps Ethernet (SysKonnect SK-9D21 Gig copper) | 219 |
| DVD (Sony DDY1621 16x DVD) | 45 |
| Floppy & 3xIDE cables, Video Card | 57 |
| OS (WindowsXP Pro OEM) | 95 |
| Database (SQL Server 2000 MSDE) | 0 |
| Shipping | 50 |
| Labor | 100 |
| **Total** | **$2,440** |

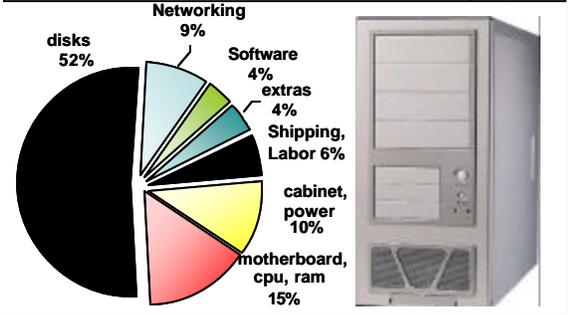

## Taxes and Shipping

The brick weights slightly under 30 pounds (14 kilograms) and can be shipped overnight for 81$. When shipping outside the NAFTA region, one must pay taxes to export/import the device – in this case the "used" device is still worth a thousand dollars, so taxes can be more than 500$. So, boarder-crossing taxes are another challenge for SneakerNet.

## Storage Brick Performance:

The Abit™ motherboard has both a dual integrated IDE controller and an on-board RAID controller. The brick has a single PCI slot and PCI 32/33 is limited to about 90 MBps. Since served data must travel over the PCI twice (once from the disk to memory and then from memory to network, brick input and output are limited to about 40 MBps in serving data to clients. The hardware's performance is shown in Table 5 and explained in the next paragraphs.

The first sequence of tests measures the local (inside the brick) performance of the IO subsystem. We used Microsoft's SQLIO test program that simulates both sequential and random IOs. In round numbers, a single disk sequentially reads files at 30MBps and sequentially writes at 40MBps – this is on the "outer" band of the disk (using 2-deep 64KB requests). On the inner bands the sequential IO rate drops to 20 MBps. On average the disks deliver about 25MBps reading and writing. These are "slow" 5200 rpm 9ms seek disks, and hence they deliver about ½ the number of random IOs that a fast (and 5x more expensive) SCSI disk can deliver. In particular, we measured 82 random reads per second and 94 random writes using 1-deep 8KB requests. When all the disks are read in parallel the read rate is an impressive 160MBps and the write rate is 84MBps.

The memory-to-memory network speed was measured with a fast client (a Dell 530 dual 2.2 GHz cpus with a server-class SysKonnect™ Gbps Ethernet card) talking to the brick via a NetGear™ GS508T Ethernet switch. The NT ttcp utility (using single threaded asynchronous 64KB requests) delivered about 90 MBps sending to the brick and 62 MB receiving from the brick. In both cases the brick was nearly cpu saturated.

**Table 5**: 3GT IO performance.

| Local SQL IO | | Read | Write |
|---|---|---|---|
| Single disk | random IOps | 82 | 94 |
| | sequential MBps | 30 | 40 |
| 7 Disk RAID0 | random IOps | 424 | 516 |
| | sequential MBps | 160 | 84 |

NT ttcp –m 1,0, <ip> -a (single threaded async 64kb)

| NT ttcp | MBps (Bytes!) | Client %cpu | Brick %cpu | Client cpb | Brick Cpb |
|---|---|---|---|---|---|
| Brick receive | 90MBps | 78% | 81% | 18 | 13 |
| Brick send | 62MBps | 32% | 89% | 12 | 21 |

| Remote SQLIO | | Read | Write |
|---|---|---|---|
| Single disk | random IOps (8KB transfers) | 80 | 89 |
| | sequential MBps (1MB transfers) | 24 | 25 |
| 7 Disk RAID0 | random IOps | 340 | 410 |
| | sequential MBps | 35 | 34 |

The SQLIO tests were then repeated on the client accessing the brick via the Ethernet. Table 5 shows the results: single disk speeds are in the 25MBps range while multi-disk sequential reads and writes saturate at about 35MBps. At that rate, it takes 11 hours to read or write a terabyte.

**Storage Brick Security:**

Viruses can be carried by foreign files (or floppy disks or CDs or FTP or HTTP.) They can contain dangerous programs. There are clear limits to what a foreign tape or floppy or CD can do to your system – but it can contain bad programs and viruses. An auto-play CD or a self-installing program opens the door wider. Virus scanners and firewalls exist to defend against such attacks.

Bringing a foreign computer inside your firewall is dangerous. A storage brick is a potential source of viruses and a potential source of attack. Defensive measures are needed against both these threats unless the subscriber completely trusts the publisher. Even if it comes from a trusted source, a "man in the middle" could ambush the BIOS or install VMware™ to ambush the software. So you cannot easily just replace the boot image and assume you own the machine (both the BIOS and the OS can spoof such operations). There are discussions of using smartcards or some form of secure-mode processor and BIOS, but those solutions are not available today.

One safe solution is to put the computer on the other side of your firewall. The goal was to reduce the phone bill, and shipping the brick does that. We do not have to eliminate the firewall if we just want to quickly and cheaply move the terabytes.

Another solution is to trust the person who sent you the system, or to trust your own internal security system to protect itself. We do not recommend that, but it is indeed what we are doing.

So, we confess that the security problem may well be a barrier to cross-company exchanging data via storage bricks. But there may be technical solutions to this if it becomes standard practice (obvious solutions like scrubbing the machine's OS, or running a virus scan are all susceptible to deceptions since the brick arrives with an unknown BIOS and operating system).

Within an organization, or among cooperating (and trusting) organizations the idea of moving bulk data via storage bricks has considerable merit. For example, these trust issues do not arise for archiving or backup/restore.

**Compression and Encryption**

Tape vendors often quote "compressed" storage capacity, claiming compression ratios of 2.5:1. So a 40GB tape is advertised as 100GB and a 4MBps data rate is quoted at 10MBps. A storage brick can play the same game. It can compress the data and so our bricks could claim to be 2.5TB rather than 1TB and the data rate could claim to be 60MBps rather than 25MBps.

A more interesting possibility is that the operating system can encrypt all the data on the disks, and so prevent the data from being disclosed to anyone not knowing the decryption key.

Compression and encryption can certainly be done with passive media, but the nice thing about storage bricks is that the compression and encryption algorithms are co-located with the data – so one is archiving or exchanging both the data and the algorithms. When the algorithms and data are separated (as with passive media like tape or CD), the algorithms may not exist at the receiver. This is especially important for long-term data preservation.

## Summary

Until we all have inexpensive end-to-end gigabit speed networks, terascale datasets will have to move over some form of sneaker net. We suspect that by the time the promised end-to-end gigabit NGI arrives, we will be moving petabyte scale datasets and so will still need a sneaker net solution. Given that, the real choice is disk, or tape, or brick. We believe that the speed, convenience, and compatibility of bricks will make them the medium of choice in the future.

Storage bricks can also be used for data interchange, for data archiving (send the whole brick to Iron Mountain), and for backup/restore.

## Acknowledgments

Thanks to Galen Hunt for helping us think through the security issues and to Leonard Chung for a modern view of CD technology.